\documentclass[traditabstract]{aa} 
\usepackage{longtable}
\usepackage{latexsym}
\usepackage{amssymb}
\usepackage{lscape}
\usepackage[]{natbib}

\usepackage{color}

\usepackage{graphicx}
\usepackage{txfonts}
\def\lsim{\mathrel{\rlap{\lower 3pt \hbox{$\sim$}} \raise 2.0pt \hbox{$<$}}}
\def\gsim{\mathrel{\rlap{\lower 3pt \hbox{$\sim$}} \raise 2.0pt \hbox{$>$}}}

   \title{Bars as seen by \textit{Herschel} \& Sloan}
   \author{Guido Consolandi \inst{1}  \and Massimo Dotti \inst{1} \and Alessandro Boselli \inst{2} \and Giuseppe Gavazzi \inst{1} 
   \and Fabio Gargiulo \inst{1} 
   }
   \authorrunning{G. Consolandi}
   \titlerunning{Bars in HRS}
   \institute{
    Dipartimento di Fisica G. Occhialini, Universit\`a di Milano-Bicocca, Piazza della Scienza 3, I-20126 Milano, Italy\\
    \email{guido.consolandi@mib.infn.it}\\
    \and
    Aix Marseille Universit\'e, CNRS, LAM (Laboratoire d'Astrophysique de Marseille), Marseille, France\\
    }
\begin{document}
\date{}

  \abstract{
  We present an observational study of the effect of bars on the  gas component and on the
  star formation properties of their host galaxies in a statistically significant sample of resolved objects, the $Herschel$ Reference Sample. The analysis of
optical and far--infrared images  allows us to identify a clear spatial correlation between
  stellar bars and the cold-gas distribution mapped by the warm dust emission.
  We find that the infrared counterparts of optically identified bars are
either  bar--like structures or dead central regions in which
  star formation is strongly suppressed. Similar morphologies are found in the distribution of 
  star formation directly traced by H$\alpha$ maps. The sizes of such optical and
  infrared structures  correlate remarkably well, hinting at a causal
connection.  In the light of previous observations and of theoretical investigations in the literature, we interpret our findings 
as further evidence of the scenario in which bars drive strong inflows toward their host nuclei: 
young bars are still in the process of	perturbing the gas and star formation clearly delineates the shape of the bars;
old bars on the contrary already removed any gas within their extents, carving  a dead region of negligible star formation.
 }
\keywords{Galaxies: structure  -- Galaxies: star formation -- Galaxies: evolution}

\maketitle


\section{Introduction}
Stellar bars are common features in disc galaxies on a broad range of stellar masses and local environments \citep[e.g.][]{jo04,sh08, bara08, nairbfrac, masters12,
skibba12, abreu12, pg15}. Because of their elongated shape, 
bars can exert a significant gravitational torque onto the host galaxy stellar and gaseous components, making these features
one of the main drivers of galactic evolution \citep[see e.g.][for recent reviews]{KK04,K13,sell14}.
In particular, the interaction between the bar and the ISM
within the bar extent results in fast inflows of gas toward the galactic center \citep{sanders76, roberts79, atha92, sakam99}. Such inflows can 
trigger nuclear bursts of star formation \citep[SF, as observationally confirmed by][]{Ho97, Martinet97, Hunt99, jo05, lauri10}, and, if the gas infall proceeds 
unimpeded, accretion episodes onto the central massive black hole \citep[if present, e.g.][]{shlo89, beren98}. 

Only recently \cite{verley07}, \cite{cheung13}, \cite{pg15} and \cite{fanali15} suggested that the prompt gas removal, if converted into stars on the short dynamical
time-scale of the galaxy nucleus, quenches any SF  in the central few kpc region of the galaxy. This scenario has strong implications
 for the evolution of the SF rate (SFR) observed in field disc galaxies as a function of their stellar mass, with the  decline of the specific SFR
(sSFR) observed in massive galaxies possibly linked to the formation of a stellar bar \citep{pg15, C16}. On the other hand,
even if bars do not remove all the gas within their extent, they are expected to perturb the gas kinematics  by pumping turbulence
in the ISM, preventing the gas from fragmenting and decreasing the central SFR  \citep[e.g.][]{reynaud98,haywood16}.

In order to test the two  above-mentioned scenarios, in this study we aim at mapping the distribution of gas in barred and unbarred galaxies.
The most direct probe would require the direct imaging of molecular and neutral atomic gas. Unfortunately, such information is available only for
a very limited sample of galaxies, and is often affected by either a too low angular resolution or a very limited field of view. 
However these problems can be overcome  because the molecular gas distribution correlates strongly with the distribution of the cold dust component \citep{boselli02}.
We take full advantage of such leverage by using the far--infrared (FIR) images from the \textit{Herschel} Reference Survey \citep[HRS, ][]{boselli10}.
We compare the {\it Herschel} data with the optical images from the Sloan Digital Sky Survey \citep[SDSS,][]{sdss}. 
We study the correlation between the occurrence of bars in optical images and of either bar-like structures
or central zones of no emission in the HRS. We further measure the extent of such optical and infrared structures and check
whether they are correlated. For galaxies showing both an optical bar and an infrared bar-related structure we link their morphology to
the star formation distribution as traced by H$\alpha$ images \citep{kc98}. 
Finally a qualitative comparison to the few available HI-maps tracing the atomic gas distribution is accomplished owing to the high resolution maps from the VIVA survey \citep{VIVA}.
\section{The \textit{Herschel} Reference Sample}
\begin{figure*}
\begin{centering}
\includegraphics[width=12.cm]{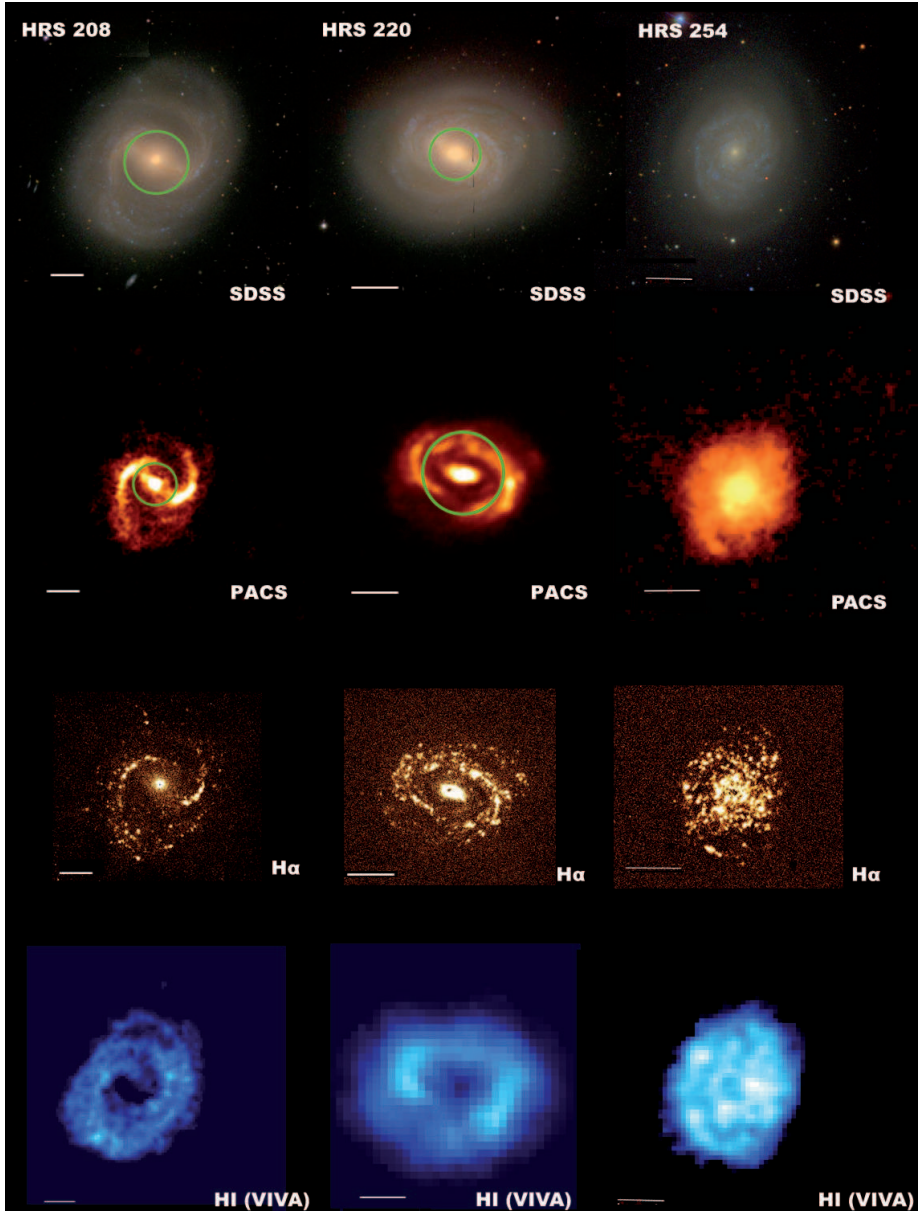}
\caption{Examples for the categories classified in this work. From left to right:  NGC 4548 (HRS-208), NGC 4579 (HRS-220), NGC 4689 (HRS-256). 
The top row shows the SDSS RGB image of the galaxy while the second row shows the corresponding PACS images. Green circles illustrate
qualitatively the circular region  used to measure the extensions of structures.
In the third row is reported the H$\alpha$ image and in the fourth the HI map from the VIVA survey.
In each frame a 1 arcminute scale is given.
}
\label{examp}
\end{centering}
\end{figure*}
The galaxies analyzed in this work have been extracted from the
\textit{Herschel} Reference Survey, a volume-limited (15$\leq$ $D$ $\leq$ 25
Mpc), $K$-band-selected sample of nearby galaxies spanning a wide range 
of morphological types, from ellipticals to dwarf irregulars, and stellar masses
(10$^8$ $\lesssim$ M$_{*} \lesssim$ 10$^{11}$ M$_{\odot}$) that has been observed in guaranteed
time with \textit{Herschel} \citep{boselli10}. 

Since the present work aims at performing a visual comparison of the ISM  and of the 
stellar morphology in the HRS galaxies, we need a sufficient spatial resolution in both the 
IR and the optical images as well as a good sensitivity and little dust obscuration in
the optical band.
For this purpose we characterize the morphological properties of the
stellar component and of the ISM using the SDSS images in the $i$-band (Cortese et
al. 2012) and the 160 $\mu$m maps obtained
with the PACS instrument (Cortese et al. 2014), respectively. At 160 $\mu$m the resolution is FWHM = 11.4 arcsec, 
while the pixel size of the reduced maps is 2.85 arcsec pixel$^{-1}$ (Cortese et al. 2014). 
This photometric band has been chosen
among those available for the whole sample galaxies (22$\mu$m from WISE,
Ciesla et al. 2014; 100-500 $\mu$m, Ciesla et al. 2012; Cortese et al. 2014)
as the best compromise in terms of sensitivity, angular resolution and dust temperature. 
At this frequency the FIR emission gives the distribution of the cold dust component, which is
a direct tracer of the molecular gas phase of the ISM (e.g. Boselli et al. 2002) from galactic to sub-kpc scales \citep{corbelli12,smith12,bolatto13,sand13}.
On the optical side, the $i$-band is only little affected by dust and is the best SDSS tracer
of the stellar mass of a galaxy, and is preferred to the $z$-band for its higher sensitivity, while
the H$\alpha$ data are taken from \citet{boselli15}. 
The SDSS, PACS and H$\alpha$ images are available on the HeDaM database (http://hedam.lam.fr/). 

Further on, we limited the analysis to the 261 late-type galaxies of the sample in order to avoid contamination from 
slow rotators \citep[namely ellipticals, which do not develop bars,][]{sell14} and from early-type disks (including S0s)
that have too little  cold gas to test the bar-related quenching process \citep{boselli14b}. 
Finally we exclude galaxies with an axis ratio lower 
than $0.4$ to avoid  a  major inclination bias in our morphology classification and measures, leaving a final subsample of 165 late-type face-on galaxies.
\section{Results}
For each galaxy we visually inspect the $i$-band SDSS images and look for the presence of an evident stellar bar.
Separately we also visually inspect the PACS images looking for a central carved region with little to no emission that, if present, is distributed 
along a bar-like component (see Fig.\ref{examp}, HRS208) or in a small nuclear region surrounded by a ring-like structure (see Fig.\ref{examp}, HRS220). 
In Fig. \ref{examp}, from left to right, we show three illustrative cases representing the infrared morphologies
possibly associated with optical bars (column one and two, HRS 208 and 220) and a normal spiral galaxy (last column, HRS 254). 
For each galaxy we give from top to bottom the 
 SDSS RGB, the 160$\mu\rm m$ , the  continuum subtracted H$\alpha$ images and the HI map from the VIVA survey \citep{VIVA} that unfortunately 
overlaps our sample only with few galaxies.\\
We find 51 barred galaxies ($\approx30\%$ of the sample) in the $i$-band, out of which
$75\%$  show in the  corresponding  160$\mu\rm m$ images an elliptical/circular area where the only emission is distributed on 
a bar- or ring-like structure. 
On the other hand, we find 63 galaxies ($\approx38\%$ of the sample) hosting the described morphologies in the 160$\mu\rm m$ images out of which 38 ($\approx65\%$) galaxies are found barred in the
corresponding optical image. 
The frequency of galaxies hosting an infrared feature that also show a corresponding optical bar, and the occurrence of optical bars showing 
an infrared feature are $\approx 65 \%$ and $\approx 75\%$, respectively.
These percentages rise to  $\approx 85 \%$ and $\approx 96\%$ if we include 16 galaxies classified as barred by other literature classifications found in the 
NASA Extragalactic Database (NED). These 
are mostly weak bars that are difficult to recognize visually and whose extent is difficult to quantify. For this reason we exclude these objects from our further analysis.
In order to quantitatively relate the region of star formation avoidance to the presence of an optical bar, we measure the size of these structures visually in the optical and 
in the 160$\mu\rm m$ and then do the same using ellipse fits to isophotes. The two approaches are useful because the eye can effectively recognize features and their extent even if somewhat
subjective, while ellipse fits are objective measures that nevertheless can be strongly affected by other structures in the galaxies.
Four of the authors (GC, MD, FG, GG) manually evaluated the extent of optical bars by measuring the radius of 
the circular region circumscribing the bar, avoiding possible HII regions at the end of it. 
On the other hand, in the 160$\mu\rm m$ images, when an infrared bar is present we measure the radius of 
the circle circumscribing the bar while, when no clear bar is discernible, we measure the inner semi-major axis of the ring-like
structure surrounding the depleted region (as depicted in Fig. \ref{examp}).
For the optical bars showing the region of avoidance in the 160$\mu\rm m$ images we also visually inspect the continuum subtracted H$\alpha$ images
finding similar morphologies and repeat the same measure.

\begin{figure}
\begin{centering}
\includegraphics[width=9.cm]{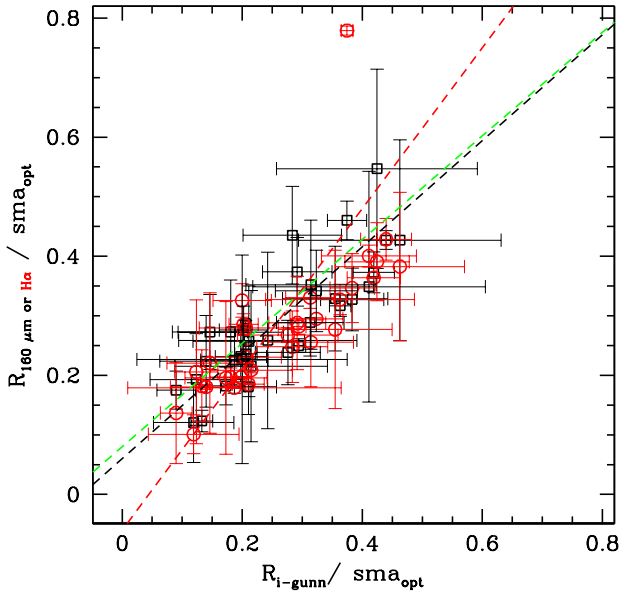}
\caption{  Comparison between the  radii of bars in the $i$-band and the radii of the 
central zone of avoidance of the 160$\mu\rm m$ (black dots) and continuum subtracted H$\alpha$ (red dots) images.
The black and red dashed lines represent the bisector regression to the $i$-band versus 160$\mu\rm m$ and $i$-band versus H$\alpha$  data, respectively.
The green dashed line is the linear fit to the $i$-band versus 160$\mu\rm m$ radii measured using $ellipse$ in IRAF for $75\%$ of the sample. 
All radii are normalised to the optical diameter of the galaxy taken from \citet{cortese12}. A comparison between the visual and automatic optical and 160$\mu\rm m$ radii is available in the
online material.
}
\label{rrr}
\end{centering}
\end{figure}

Using IRAF\footnote{IRAF (Image Reduction and Analysis Facility) is a software for the reduction and analysis of astronomical data.} task $ellipse$, ellipticity ($\epsilon$) and position angle (P.A.) radial profiles of the isophotes of each sample galaxy 
in each considered band.
In optical broad-bands, it is well tested that the radius at which there is a peak in the ellipticity profile 
and a related plateau in the P.A. profile is a good proxy for   the extension of the bar \citep{jo04,lauri10,C16b}.
Following this procedure we extract a radius of the bar in the $i$-band for each galaxy and, 
we deduce the  bars strength following \citep{lauri07}  from the peak of the $\epsilon$  profile in the $i$-band. We find that $\approx 95\%$ of galaxies that we classified as barred 
harbor strong bars ($\epsilon>0.4$).
Although this quantitative method has not been applied to far-IR data previously, ellipse fits can nevertheless be derived for the 160$\mu\rm m$ images and the $\epsilon$ and P.A. profiles
examined for a bar signature. Since we are trying to measure a region of decreased emission possibly surrounded by a ring-like
emitting structure, we also extract a radial surface brightness profile from concentric elliptical apertures centered on the galaxy and ellipticity fixed to the outer infrared isophotes.
The derived surface brightness profile therefore has a relative maximum in correspondence of the ring-like structure.
In the cases where no infrared bar is discernible, the radius at which this occurs is a good proxy of the extension of the non emitting region.\\
In the 160$\mu\rm m$ images, this method succeeds at extracting the radius of the non emitting region or of the bar in $75\%$ of the barred galaxies.
Because of the irregular and clumpy distribution of light at 160$\mu\rm m$ the fit of the isophotes does not converge in $25\%$ of the galaxies.
Therefore, in order to preserve the statistics of our already limited sample, we plot in Fig. \ref{rrr} the radius obtained averaging the measures of the optical bars 
made by the authors versus those from the 160$\mu\rm m$ data (black empty dots) and those from the continuum subtracted H$\alpha$ images (red empty dots). 
All radii are normalized to the  $i$-band $25^{th}$ mag isophote radius of the galaxy, taken from \citet{cortese12}, and errors are evaluated from the standard deviation of our measurements.
The black and red dashed lines indicate the bisector fit \citep{bisfit} to the data respectively for the 160$\mu\rm m$ (slope $\sim0.89\pm0.11$) and the H$\alpha$ data (slope $\sim1.35\pm0.08$).
The slope of the fit relative to the optical versus H$\alpha$ data is strongly influenced by the extremely deviant point (associated to HRS 322) visible in Fig. \ref{rrr}. This outlier is characterized
by a very small error, as all the authors consistently measured the same radius with very little scatter. We stress, however, that 
the semi-major axis of this galaxy in the H$\alpha$
distribution is perpendicular to the optical bar (see the optical and H$\alpha$ image in the online material) thus the important discrepancy is mostly due to projection effects. If we exclude this point
from the fit, the slope becomes $0.79\pm0.11$.
Finally, in green, we plot the best linear fit of the comparison of optical versus 160$\mu\rm m$ radii measured with IRAF (slope $\sim0.87\pm0.10$).
All fits show a strong consistency between them even when evaluated with independent methods.
To further check a possible bias due to inclination, we derived the same fits for a subsample of galaxies with axis ratio greater than 0.7 ($\approx40\%$ of the sample), 
finding fully consistent results.
\section{Discussion and conclusions}
The study and comparison of frequencies of occurrence of bar related features in the optical and FIR, as traced by the stellar continuum and
by warm dust emission, respectively, results in a fraction of galaxies hosting an optical bar of $\sim30\%$ while
a  zone of avoidance with or without an infrared bar is found in $38\%$ of the 160$\mu\rm m$ images. The percentages
of common occurrence suggests that FIR images are an effective way of identifying the presence of a bar in a galaxy.

For the galaxies hosting both an optical bar and a central zone of avoidance in the 160$\mu\rm m$ images,
we measured the angular size of both structures with independent methods, finding a good correspondence.  
First, we measured the extent of bars in optical images,
while in the FIR images we measured the extent of the bar-like structure, if present, or of the inner semi-major axis of the
ring-like structure. In the latter case,
we stress that the projected angular sizes of the optical bar and the radius of non-emitting zone
may differ significantly\footnote{Up to a factor of $\approx 2.5$
for the maximum inclination of $B/A=0.4$ allowed in our sample.} depending on the bar orientation. 
In $75\%$ of the barred galaxies we successfully ran the IRAF task $ellipse$ to objectively measure the extent of these structures 
in both the  $i$-band and 160$\mu\rm m$ images, using the derived ellipticity, P.A. and surface brightness profiles.
The goodness of the correlation strongly hints at a physical connection between
the presence of an optical strong bar and a gas-depleted/quenched region where little SF is still possible. Only 
in the very center (where the bar conveys the gas originally within its reach) or along the bar is SF found. 
Such an effect is consistent with what we see in the continuum subtracted H$\alpha$ images of the sample. 
SF is indeed distributed mainly in the nuclear region of galaxies 
and/or along the bar (consistent with \cite{verley07}; see  Fig. \ref{examp}) and shows a morphology similar to the one observed in the FIR.

We conclude that the FIR morphologies are similar to the H$\alpha$ morphologies \citep[consistent with][]{verley07} and that both are 
consistent with bar-driven inflows of gas inside the corotation radius as predicted by simulations \citep{sanders76,atha92}. Fig. \ref{examp} qualitatively
show that also the HI emission, when avilable, show similar mophologies.
The impact on the cold gas component, as derived from the FIR, is consistent to what has been observed in few galaxies \citep{sakam99} and 
affects the star formation of barred galaxies \citep{verley07,pg15}: 
as soon as a bar starts growing, the gas is initially perturbed and compressed along the bar, where it forms stars while gradually losing its angular momentum; as the time goes by, the gas is swept by the bar into sub-kpc 
scales, leaving a  gas-depleted and SF quenched region of the size of the bar itself, with or without a central knot of SF depending on the consumption timescale of the originally infalling gas.
\begin{acknowledgements}
We thank the anonymous Referee and the Editor, Francoise Combes, for their constructive criticism.
This research has been partly financed by the French national program PNCG
, and it has made use of the GOLDmine database (Gavazzi et
al. 2003, 2014b) and SDSS Web site \emph{http://www.sdss.org/}. \\
\end{acknowledgements}


\begin{thebibliography}{}
\bibitem[Athanassoula(1992)]{atha92} Athanassoula, E.\ 1992, \mnras, 259, 345 
\bibitem[Barazza et al.(2008)]{bara08} Barazza, F.~D., Jogee, S., \& Marinova, I.\ 2008, \apj, 675, 1194 
\bibitem[Berentzen et al.(1998)]{beren98}  Berentzen, I., Heller,  C. H., Shlosman, I. \& Fricke, K. 1998, MNRAS, 300, 49
\bibitem[Bolatto et al.(2013)]{bolatto13} Bolatto, A.~D., Wolfire, M., \& Leroy, A.~K.\ 2013, \araa, 51, 207 
\bibitem[Boselli et al.(2002)]{boselli02} Boselli A., Lequeux J., \& Gavazzi G.\ 2002, \aap, 384, 33 
\bibitem[Boselli et al.(2010)]{boselli10} Boselli A., Eales S., Cortese L., et al.\ 2010, \pasp, 122, 261
\bibitem[Boselli et al.(2014)]{boselli14b} Boselli, A., Cortese, L., Boquien, M., et al.\ 2014, \aap, 564, A66 
\bibitem[Boselli et al.(2015)]{boselli15} Boselli, A., Fossati, M., Gavazzi, G., et al.\ 2015, \aap, 579, A102
\bibitem[Cheung et al.(2013)]{cheung13} Cheung, E., Athanassoula, E., Masters, K.~L., et al.\ 2013, \apj, 779, 162 
\bibitem[Chung et al.(2009)]{VIVA} Chung, A., van Gorkom, J.~H., Kenney, J.~D.~P., Crowl, H., \& Vollmer, B.\ 2009, \aj, 138, 1741 
\bibitem[Ciesla et al.(2012)]{ciesla12} Ciesla L., Boselli A., Smith M.~W.~L., et al.\ 2012, \aap, 543, A161 
\bibitem[Ciesla et al.(2014)]{ciesla14} Ciesla L., Boquien M., Boselli A., et al.\ 2014, \aap, 565, A128 
\bibitem[Consolandi et al.(2016)]{C16}  Consolandi, G., Gavazzi, G., Fumagalli, M., Dotti, M., \& Fossati, M.,\ 2016, \aap, 591, A38
\bibitem[Consolandi(2016)]{C16b} Consolandi, G.\ 2016, arXiv:1607.05563 
\bibitem[Corbelli et al.(2012)]{corbelli12} Corbelli, E., Bianchi, S., Cortese, L., et al.\ 2012, \aap, 542, A32 
\bibitem[Cortese et al.(2012)]{cortese12} Cortese L., Boissier S.,Boselli A., et al.\ 2012, \aap, 544, A101 
\bibitem[Cortese et al.(2014)]{cortese14} Cortese L., Fritz J., Bianchi S., et al.\ 2014, \mnras, 440, 942 
\bibitem[Fanali et al.(2015)]{fanali15} Fanali R., Dotti M., Fiacconi D., \& Haardt F.\ 2015, \mnras, 454, 3641 
\bibitem[Gavazzi et al.(2003)]{pg03} Gavazzi, G., Boselli, A., Donati, A., Franzetti, P., \& Scodeggio, M.\ 2003, \aap, 400, 451 
\bibitem[Gavazzi et al.(2014)]{pg14b} Gavazzi, G., Franzetti, P., \& Boselli, A.\ 2014, arXiv:1401.8123 
\bibitem[Gavazzi et al.(2015)]{pg15} Gavazzi, G., Consolandi, G., Dotti, M., et al.\ 2015, \aap, 580, A116
\bibitem[Haywood et al.(2016)]{haywood16} Haywood M., Lehnert M.D., Di Matteo P., et al.\ 2016, \aap, 589, A66
\bibitem[Ho et al. (1997)]{Ho97} {Ho} L.~C., {Filippenko} A.~V., {Sargent} W.~L.~W., 1997, \apj, 487, 591
\bibitem[{{Hunt} \& {Malkan}(1999)}]{Hunt99}{Hunt} L.~K., {Malkan} M.~A., 1999, \apj, 516, 660
\bibitem[Isobe et al.(1990)]{bisfit} Isobe, T., Feigelson, E.~D., Akritas, M.~G., \& Babu, G.~J.\ 1990, \apj, 364, 104
\bibitem[Jogee et al.(2004)]{jo04} Jogee, S., Barazza, F.~D., Rix, H.-W., et al.\ 2004, \apjl, 615, L105 
\bibitem[Jogee et al.(2005)]{jo05} Jogee, S., Scoville, N., \&  Kenney, J.~D.~P.\ 2005, \apj, 630, 837
\bibitem[Kennicutt(1998)]{kc98} Kennicutt, R.~C., Jr.\ 1998, \araa, 36, 189 
\bibitem[Kormendy \& Kennicutt(2004)]{KK04} Kormendy, J., \& Kennicutt, R.~C., Jr.\ 2004, \araa, 42, 603 
\bibitem[Kormendy(2013)]{K13} Kormendy, J.\ 2013, Secular Evolution of Galaxies, 1 
\bibitem[Laurikainen et al.(2007)]{lauri07} Laurikainen, E., Salo, H., Buta, R., \& Knapen, J.~H.\ 2007, \mnras, 381, 401 
\bibitem[Laurikainen et al.(2010)]{lauri10} Laurikainen, E., Salo, H., Buta, R., Knapen, J.~H., \& Comer{\'o}n, S.\ 2010, \mnras, 405, 1089
\bibitem[Martinet \& Friedli (1997)]{Martinet97} Martinet L., \& Friedli D., 1997, \aap,323,363
\bibitem[Masters et al.(2012)]{masters12} Masters, K.~L., Nichol, R.~C., Haynes, M.~P., et al.\ 2012, \mnras, 424, 2180 
\bibitem[M{\'e}ndez-Abreu et al.(2012)]{abreu12} M{\'e}ndez-Abreu, J., S{\'a}nchez-Janssen, R., Aguerri, J.~A.~L., Corsini, E.~M., \& Zarattini, S.\ 2012, \apjl, 761, L6 
\bibitem[Nair \& Abraham(2010)]{nairbfrac} Nair, P.~B., \& Abraham, R.~G.\ 2010, \apjl, 714, L260 
\bibitem[Reynaud \& Downes(1998)]{reynaud98} Reynaud, D., \& Downes, D.\ 1998, \aap, 337, 671 
\bibitem[Roberts et al.(1979)]{roberts79} Roberts W.W., Jr., Huntley J.~M., \& van Albada G.D.\ 1979, \apj, 233, 67 
\bibitem[Sanders \& Huntley(1976)]{sanders76}Sanders, R. H. \&   Huntley, J. M. 1976, ApJ, 209, 53
\bibitem[Sandstrom et al.(2013)]{sand13} Sandstrom, K.~M., Leroy, A.~K., Walter, F., et al.\ 2013, \apj, 777, 5 
\bibitem[Sakamoto et al.(1999)]{sakam99} Sakamoto, K., Okumura, S.~K., Ishizuki, S., \& Scoville, N.~Z.\ 1999, \apj, 525, 691 
\bibitem[Sheth et al.(2008)]{sh08} Sheth, K., Elmegreen, D.~M., Elmegreen, B.~G., et al.\ 2008, \apj, 675, 1141 
\bibitem[Sellwood(2014)]{sell14} Sellwood, J.~A.\ 2014, Reviews of Modern Physics, 86, 1 
\bibitem[Shlosman et al. (1989)]{shlo89} Shlosman I., Frank J., Begelman M.C., 1989, Nature, 338, 45
\bibitem[Smith et al.(2012)]{smith12} Smith, M.~W.~L., Eales, S.~A., Gomez, H.~L., et al.\ 2012, \apj, 756, 40 
\bibitem[Skibba et al.(2012)]{skibba12} Skibba, R.~A., Masters, K.~L., Nichol, R.~C., et al.\ 2012, \mnras, 423, 1485
\bibitem[Verley et al.(2007)]{verley07} Verley, S., Combes, F., Verdes-Montenegro, L., Bergond, G., \& Leon, S.\ 2007, \aap, 474, 43 
\bibitem[York et al.(2000)]{sdss} York, D.~G., Adelman, J., Anderson, J.~E., Jr., et al.\ 2000, \aj, 120, 1579 
\end{thebibliography}
\end{document}